\documentclass[%
 reprint, 
 amsmath,amssymb,
 aps,
longbibliography
]{revtex4-1}

\usepackage{enumitem}
\usepackage{graphicx}
\usepackage{qcircuit}

\usepackage{bm}

\usepackage{xcolor}
\usepackage[colorlinks=true,citecolor=blue,linkcolor=magenta]{hyperref}

\begin{document}

\title{Quantum Chaos on Complexity Geometry}

\author{Bin Yan}
\affiliation{Center for Nonlinear Studies, Los Alamos National Laboratory, Los Alamos, New Mexico 87545, USA}
\affiliation{Theoretical Division, Los Alamos National Laboratory, Los Alamos, New Mexico 87545, USA}
\author{Wissam Chemissany}
\affiliation{Institute for Quantum Information and Matter, California Institute of Technology, Pasadena, California 91125, USA.}

\date{\today}

\begin{abstract}
This article tackles a fundamental long-standing problem in quantum chaos, namely, whether quantum chaotic systems can exhibit sensitivity to initial conditions, in a form that directly generalizes the notion of classical chaos in phase space. We develop a linear response theory for complexity, and demonstrate that the complexity can exhibit exponential sensitivity in response to perturbations of initial conditions for chaotic systems. Two immediate significant results follows: i) the complexity linear response matrix gives rise to a spectrum that fully recovers the Lyapunov exponents in the classical limit, and ii) the linear response of complexity is given by the out-of-time order correlators.
\end{abstract}

\maketitle

The existence of quantum chaos has been questioned for a long time~\cite{Haake2006-ic,Berry2006-gh,cvitanovic2016chaos}. Quantum dynamic is fundamentally unitary and inner-product preserving; hence, in terms of state overlaps, quantum evolution does not exhibit sensitive dependence on initial conditions -- the telltale signature of classical chaos. 

On the other hand, it has been well-known that quantum systems with chaotic and regular classical counterparts display distinct characteristics. Eventually, various diagnostics have been employed to exhibit chaotic behavior in quantum systems. Conventional approaches utilize the spectral properties of chaotic Hamiltonians, e.g., random matrix~\cite{Kota2014-cg} or periodic orbit theory~\cite{Gutzwiller1990-sx}. Recent developments address the problem in the time domain. These include the Loschmidt echo~\cite{Jalabert2001-kn,Gorin2006-ro}, entropy production in open systems~\cite{Zurek1994-om,Zurek2006-vj}, or out-of-time order correlator (OTOC)~\cite{Kitaev2015,Larkin1969,Maldacena2016-mb}, which is responsible for the revival of interest in quantum chaos (See Ref.~\cite{Swingle2018-xe} for a review). 
However, none of these approaches were able to establish a direct analog to classical chaos in terms of the sensitivity to initial conditions. Consequently, some even suggest to call this field quantum chaology~\cite{Berry2006-gh}.

One may argue that quantum wavefunction is not a good analogue of the classical state. The latter is a single point in phase space, while the former corresponds to a probability density, whose classical counterpart should be the Liouville's distribution. The overlap between two Liouville's density of states under classical dynamics remains constant as well~\cite{cvitanovic2016chaos}. This suggests that the overlap, or precisely inner-product based metrics, might not be a good measure for quantifying the difference between two quantum ``trajectories''. 

\begin{figure}[t!]
    \centering
    \includegraphics[width=\columnwidth]{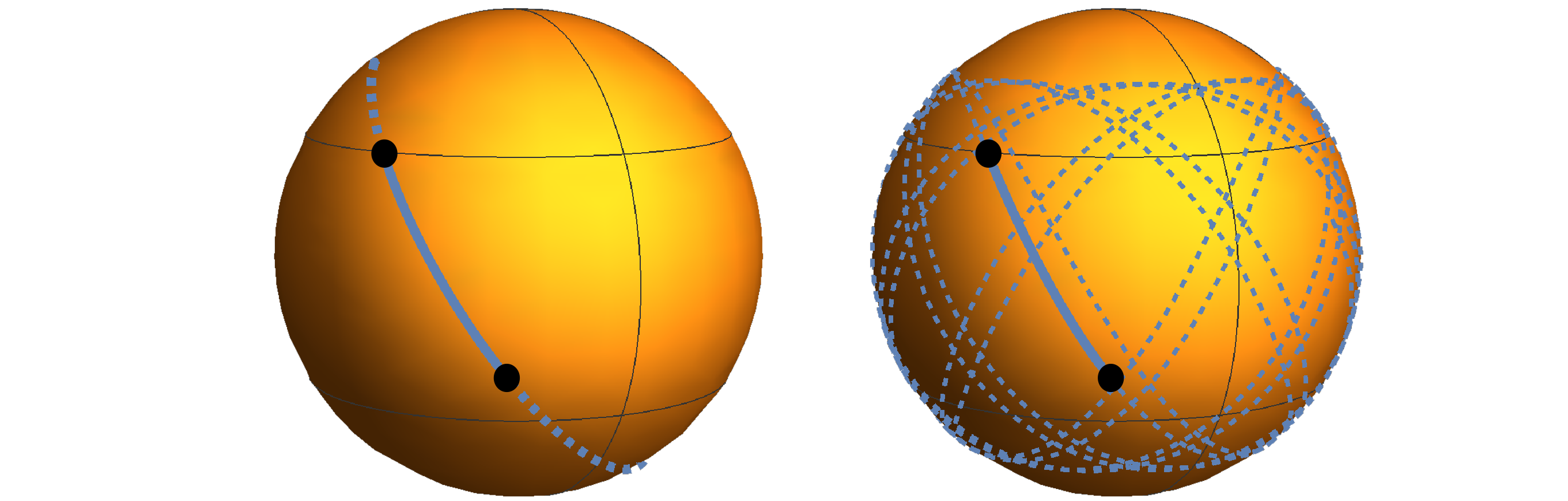}
    \caption{Complexity metric can induce a different geometry than the usual inner-product metric: The distance between any two states (black dots) on the Bloch sphere is given by the length of the shortest geodesic (solid curves). The geodesic under the inner-product metric is a great circle (dashed curve on the left). The closed geodesic under the complexity metric~[24] can be highly complex (dashed curve on the right).}
    \label{figure}
\end{figure}

More recently, a novel distance measure, known as the relative complexity between quantum states, and its implications to the problem of quantum chaos, have attracted considerable attention~\cite{Jefferson2017-wf,brown2018second,brown2017quantum,Chapman2018-sy,Khan2018-ow,Magan2018-ua,Bueno2019-zh}. The notion of relative complexity between states was adopted from the circuit complexity of unitaries, which has an attractive geometric formalism developed by Nielsen \emph{et al.}~\cite{nielsen2005geometric,Dowling2008-pu,nielsen2006quantum}. Figure~\ref{figure} illustrates how the complexity metric changes the geometry of the wavefunction manifold \footnote{The complexity metric used in the figure is induced by the cost function $\mathcal{F}=\int_0^1\ ds \sqrt{|Y^x(s)|^2+|Y^y(s)|^2+1.5|Y^z(s)|^2}$ for the three Pauli matrices. The usual inner-product metric corresponds to a cost function with equal weight for all Pauli matrices.}, even for a single qubit~\cite{Brown2019-cn}. The complexity distance between two time-dependent wavefunctions, which are initially close to each other, can grow exponentially in time~\cite{Bueno2019-zh}. Note that the two initially nearby quantum states are generated by small perturbations; hence, the complexity truly detects the sensitive dependence of a chaotic quantum system to its initial condition.

Despite these developments, it is still not clear whether the complexity theory can fully recover classical chaos in the proper limit. As a framework that claims to generalize classical chaos to the quantum regime, it should give predictions that coincide with the classical one in the classical limit. Another important problem yet to be explored is the relation between complexity and other measures for quantum chaos, such as the OTOCs. 

With all these questions in mind, we develop a linear response theory for complexity. A \emph{response matrix} is introduced to characterize the fine structure of the complexity in response to initial perturbations. Significantly, the response matrix gives rise to the full Lyapunov spectrum in the classical limit. Moreover, we demonstrate that the complexity response is essentially given by the OTOCs of the corresponding operators. This intrinsic connection provides an alternative and yet intuitive way of justifying OTOC as a diagnostic for chaos.

\vspace{5pt}
\emph{Complexity---}
Roughly speaking, the circuit complexity for a target unitary $U_T$ is defined as the minimum number of local quantum gates required to implement it. This leads to a geometric formalism put forward by Nielsen \emph{et al.}~\cite{nielsen2005geometric,Dowling2008-pu,nielsen2006quantum}. The unitary $U_T$ can be simulated through a parameterized Hamiltonian evolution
\begin{equation}\label{eq:path}
    U_T = e^{-i\int_0^1 d\sigma\  \sum_I Y^I(\sigma)M_I}.
\end{equation}
Here $\{M_I\}$ is the set of generators that represents the accessible operations (analogous to elementary gates) for the physical system under consideration.
The choice of the path $\{Y^I(\sigma)\}$ is in general not unique. One can define a suitable cost function $\mathcal{F}[Y(\sigma)]$, from which an optimal path with the minimum cost can be selected. We will restrict ourselves to the commonly used cost function
\begin{equation}\label{eq:cost}
    \mathcal{F}[Y(\sigma)]\equiv \int_0^1 d\sigma \sqrt{\sum_I |Y^I(\sigma)|^2}.
\end{equation}
The complexity of the target unitary is then defined as the least cost along the optimal path, i.e.,
\begin{equation}\label{eq:ucomplexity}
    \mathcal{C}^u\left[U_T\right] \equiv \min_{Y} \mathcal{F}[Y(\sigma)].
\end{equation}
The complexity of unitary can be adopted to define the complexity of a target wavefunction $|\Psi_T\rangle$, with respect to a given reference state $|\Psi_R\rangle$. Namely, it is defined as the least complexity for all unitaries that transforms the reference state to the target state,
\begin{equation}\label{eq:scomplexity}
    \mathcal{C}^s\left[|\Psi_T\rangle,|\Psi_R\rangle\right] \equiv \min_U \mathcal{C}^u\left[U\right],\ \ U|\Psi_R\rangle=|\Psi_T\rangle.
\end{equation}

The above state complexity defines a property characteristic for the target wavefunction, and is dependent on the choice of the reference state. Here, we treat the target and reference states on the same footing, and interpret the state complexity as a relation between two wavefunctions. In particular, by including the Hermitian conjugate of each operator in the generator set, the state complexity becomes a symmetric functional, constituting a metric between any two wavefunctions~\cite{brown2018second}. 

\vspace{5pt}
\emph{Linear response---}
The complexity itself cannot directly distinguish whether a unitary evolution is chaotic or regular, i.e., a unitary generated by a regular Hamiltonian for a long time can have a larger complexity than the one generated by a chaotic Hamiltonian for a short period of time~\cite{Ali2020-dk}. To study the behavior of the chaotic dynamics, one ought to look at how it responses to perturbations.

Previous studies show that, for chaotic systems, when a small perturbation is applied to the initial state, the relative complexity between the perturbed and unperturbed state can grow exponentially in time~\cite{Bueno2019-zh}. Note that the overlap between these two wavefunctions does not change in time; hence, usual distance measures based on inner-product cannot capture the sensitivity of dynamics to initial states, which is one of the key reasons for questions concerning the existence of quantum chaos~\cite{Haake2006-ic,Berry2006-gh,cvitanovic2016chaos}.

To extract the fine structure of the complexity responses to initial perturbations, we introduce the notion of \emph{partial complexity}, which, in contrast to the full complexity, qualifies only the complexity for a given generator. Denote $Y_c$ the optimal path that minimizes the cost function (\ref{eq:cost}). The partial complexity for generator $M_L$ is evaluated for the $M_L$ component of the optimal path $Y_c$, i.e.,
\begin{equation}\label{eq:pcomplexity}
    \mathcal{C}_L \equiv \int_0^1 d\sigma \ Y_c^L(\sigma).
\end{equation}
It is worth emphasizing that the optimal path $Y_c$ in the above integral is determined by minimizing the total cost function (\ref{eq:cost}). Depending on whether the optimal path is determined for the unitary or state complexity, the partial complexity can be defined accordingly for both cases. In the following, we use superscripts $u$ and $s$ to label the unitary and state version of the partial complexity, respectively.

We are now ready to study the partial (state) complexity in response to small perturbations. For two wavefunctions that initially deviate from each other by a perturbation generated through the operator $M_K$,
\begin{equation}\label{eq:initial}
\begin{aligned}
   |\Psi_1(t)\rangle = e^{-iHt} |\Psi\rangle, \ \ |\Psi_2(t)\rangle = e^{-iHt} e^{i\epsilon M_K} |\Psi\rangle,
\end{aligned}
\end{equation}
the linear response of the partial complexity for operator $M_L$, with respect to $M_K$, is defined as
\begin{equation}\label{eq:sresponse}
   R^s_{KL} \equiv \frac{\partial}{\partial \epsilon} \mathcal{C}_{L}\left[|\Psi_1(t)\rangle,|\Psi_2(t)\rangle\right]|_{\epsilon=0}.
\end{equation}

Note that computing the relative state complexity for the time dependent wavefunctions (\ref{eq:initial}) involves a double-optimization, i.e., minimizing the cost function for all possible unitaries that connects these two states. There is a particular unitary among all of them that is of interest on its own, namely,
\begin{equation}\label{eq:targetu}
    U_\epsilon(M_K) = e^{-iHt}e^{i\epsilon M_K}e^{iHt}.
\end{equation}
This is the only unitary that works for any initial state $|\Psi\rangle$. Its unitary complexity tells of information of the dynamics alone, independent of the system state. Thus, we further introduce a unitary complexity version of the linear response,
\begin{equation}\label{eq:uresponse}
   R^u_{KL} \equiv \frac{\partial}{\partial \epsilon} \mathcal{C}_{L}\left[ U_\epsilon(M_K) \right]|_{\epsilon=0}.
\end{equation}

Up to this point, we have laid down the general framework to compute complexity in response to initial perturbations. The full information is contained in the linear response matrix
\begin{equation}
    \hat{L}\equiv\hat{R}^\dag\hat{R},
\end{equation}
where the matrix $\hat{R}$ is defined through Eq.~(\ref{eq:sresponse}) or (\ref{eq:uresponse}). In the following, we present two immediate significant results of this framework: i) the linear response theory for the state complexity generalizes classical chaos in phase space. More precisely, the eigenvalues of the response matrix $\hat{L}$ gives rise to the Lyapunov spectrum in the classical limit, and ii) the linear response for the unitary complexity is intrinsically related to the out-of-time order correlator.

\vspace{5pt}
\emph{Classical limit---} We shall focus on the time scale before the Ehrenfest time, when the quantum dynamics is reduced to the classical one. This time scale is also known as the scrambling time~\cite{Maldacena2016-mb}, which is one of the primary focuses in quantum chaos. The initial states are assumed to have localized Wigner representations.

In connection to the phase space structure, the set of generators for computing complexity is chosen as the Heisenberg algebra $\{p_i,q_i,i\hbar \mathbb{I}\}_{i=1,...,N}$, where $N$ is the dimension of the phase space; The cost function takes the form (\ref{eq:cost}) for all elements in the Heisenberg algebra, except for the identity operator, which we assign a zero cost. Since the identity generates overall phases to the wavefunction, this is equivalent to say that we treat two wavefunctions which differ by a global phase as the same physical state. This convention is repeatedly used in the following discussions. The Heisenberg algebra does not generate the full unitary group, therefore is not complete to compute the complexity for arbitrary Hamiltonian. However, as will be seen in the following, in the classical limit the wavefunction evolution corresponds to a coordinate shift in phase space. In this case using Heisenberg algebra is sufficient. Alternatively, one can enlarge the algebra and assign high cost for the additional generators, such that the complexity can be computed for any Hamiltonian, but the early time behavior in the classical limit is not altered.

Denote $\bm{x}=\{x^{k}\}_{k=1,...,2N}$ the phase space coordinates, and $W_1(\bm{x},t=0)$ the Wigner function corresponding to the initial state $|\Psi_1(t=0)\rangle$. Then the perturbed wavefunction $|\Psi_2(t=0)\rangle = e^{i\epsilon x^{\bar{k}}}|\Psi_1(t=0)\rangle$ has a Winger functions that is obtained by shifting $W_1(\bm{x},t=0)$ along the direction $\hat{\bm{e}}_k$, i.e.,
\begin{equation}
    W_2(\bm{x},t=0) = W_1(\bm{x}+\epsilon \hat{\bm{e}}_k ,t=0).
\end{equation}
Here $x^k$ and $x^{\bar{k}}$ are conjugate pairs.

In the classical limit (before the Ehrenfest time), the equation of motion for the Wigner function reduces to the Liouville equation. According to the Liouville theorem, the density does not change along the trajectory, therefore,
\begin{equation}
    W_1(\bm{x},t)=W_1(\tilde{\bm{x}}_{-t}(\bm{x}),t=0),
\end{equation}
where $\tilde{\bm{x}}_{t}(\bm{x})$ is the time dependent trajectory of the initial point $\bm{x}$. Similarly,
\begin{equation}
    W_2(\bm{x},t) = W_1(\tilde{\bm{x}}_{-t}(\bm{x}+\epsilon \hat{\bm{e}}_k),t=0).
\end{equation}
To compute the complexity, note that the unitary evolution that transforms $W_1(\bm{x},t)$ to $W_2(\bm{x},t)$, must effectively generate a shift $\tilde{\bm{x}}_{-t}(\bm{x}) \rightarrow \tilde{\bm{x}}_{-t}(\bm{x}+\epsilon \hat{\bm{e}}_k)$. Since $\tilde{\bm{x}}_t$ is the classical trajectory, the infinitesimal shift, to the first order of $\epsilon$, can be expanded as, 
\begin{equation}\label{JacM}
    \tilde{\bm{x}}_{-t}(\bm{x}+\epsilon \hat{\bm{e}}_k) = \tilde{\bm{x}}_{-t}(\bm{x}) + \epsilon \frac{\partial \tilde{x}^i_{-t}}{\partial x^k}\hat{\bm{e}}_i.
\end{equation}
The classical Jacobian matrix $\partial \tilde{x}^i_{-t}/\partial x^k$ in general depends on initial position $\bm{x}$, especially for non-fully chaotic systems, whose Lyapunov exponents depend on the regime of the phase space. However, we always start from a sufficiently localized point-like Wigner function, and work with the early time of the dynamics, so that the $\bm{x}$-dependence is eliminated. Thus, the path protocol $Y(\sigma)$ must generate a shift of the Wigner function in the phase space given by the second term in Eq.~(\ref{JacM}). The complexity is then the minimum length of the path subject to this boundary condition. Subject to the cost function~(\ref{eq:cost}), the optimal path $Y_c$ is precisely given by 
\begin{equation}
    Y_c^i(\sigma)=\epsilon \frac{\partial \tilde{x}^i_{-t}}{\partial x^k}\sigma,\ \sigma \in [0,1].
\end{equation}
The linear response (\ref{eq:sresponse}) is then identified as the classical Jacobian matrix, i.e.,
\begin{equation}
    R^s_{\bar{k}i}=\frac{\partial \tilde{x}^i_{-t}}{\partial x^k}.
\end{equation}
Denote $\{s_i(t)\}$ the time-dependent eigenvalues of the linear response matrix $\hat{L}$. In analog to the classical case, the quantum Lyapunov spectrum can be extracted as
\begin{equation}\label{eq:Lspectrum}
    \lambda_i \equiv \lim_{t\rightarrow \infty} \frac{1}{2t} \ln s_i(t).
\end{equation}
We thus arrive at the conclusion that, in the classical limit, the quantum Lyapunov spectrum given by the complexity linear response fully reduces to the classical one.
It is worth emphasizing that the classical Lyapunov spectrum in general depends on the initial position in phase space. This fact is reflected in the above quantum generalization, since the state complexity depends on the initial wave function as well. In contrast, as will be discussed in the following, the linear response matrix for the unitary complexity, and consequently the OTOCs, cannot capture this initial condition dependence.

\vspace{5pt}
\emph{Complexity and OTOC---} To compute the linear response for the unitary partial complexity (\ref{eq:uresponse}), we first induce the correct boundary condition for the path. Expanding both the parameterized path (\ref{eq:path}) and the target unitary (\ref{eq:targetu}) to first order of $\epsilon$ gives
\begin{equation}
    \sum_I \int_0^1 d\sigma\ Y^I(\sigma)M_I = \epsilon M_K(t).
\end{equation}
For the particular metric we used in the cost function (\ref{eq:cost}) and infinitesimal $\epsilon$, the optimal path $Y_c$ that minimize the cost function corresponds to a straight line in the vector space of the generators, namely, $Y^I(\sigma)=Y^I_c\sigma$, $\sigma\in [0,1]$, the integral of which gives rise to the partial complexity $C_I = Y^I_c$, which satisfies
\begin{equation}
  \sum_I C^I M_I = \epsilon M_K(t).
\end{equation}
To further extract the full matrix of the linear response, we compute the commutator on both side of the above equation with the generator $M_L$, and do a partial derivative in $\epsilon$. This gives
\begin{equation}
    \sum_I R^u_{KI} [M_I,M_L]=[M_K(t),M_L].
\end{equation}
Denote $\hat{T}_{IJ}\equiv [M_I,M_J]$ as a transfer matrix, and $\hat{O}_{IJ}(t)\equiv [M_I(t),M_J]$. The above equation has a compact form $\hat{R}^u\hat{T}=\hat{O}$.

Averaging over a quantum state $|\psi\rangle$ gives
\begin{equation}\label{eq:COTOC}
    \langle\psi| \hat{O}^\dag \hat{O}|\psi\rangle = \langle\psi| \hat{T}^\dag  \hat{L}^u \hat{T} |\psi\rangle,
\end{equation}
where $\hat{L}^u\equiv (\hat{R}^u)^\dag\hat{R}^u$ is the linear response matrix for the unitary complexity.
The left hand side of the above equation contains the full information of the OTOCs for the given set of operators \footnote{OTOC in the form of the commutator square is defined as $\langle |[W(t),V]|^2\rangle$. The Heisenberg evolution of the operator $W(t)$ results in a growth of the OTOC.}. It has been proposed~\cite{Gharibyan2019-sp} as a definition for the quantum Lyapunov spectrum, when the generators are chosen to form a Heisenberg algebra. In this case, the transfer matrix takes a symplectic form
\begin{equation}
    i\hat{T}/\hbar=\begin{pmatrix}0 & -\mathbb{I}_n \\ \mathbb{I}_n & 0\end{pmatrix}.
\end{equation} 

Note that the response matrix $\hat{L}^u$ in the unitary complexity case defines a spectrum as well. However, since the target unitary (\ref{eq:targetu}) is independent of the initial state, the corresponding spectrum cannot reflect the initial position dependence of the classical Lyapunov spectrum in phase space, in contrast to the case of state complexity.

\vspace{5pt}
\emph{Model Study---} To illustrate the complexity linear responses, we present a study of the inverted harmonic oscillator (IHO), whose Hamiltonian reads
\begin{equation} 
H=\frac{1}{2}\hat{p}^{2} - \frac{1}{2}\Omega^{2} \hat{x}^{2}.
\end{equation}
The IHO is an archetype model for chaos with physical relevance~\cite{Blume-Kohout2003-ww,Magan2018-ua}, and has been recently studied in the context of quantum chaos~\cite{Morita2019-de,ali2020chaos,yan2019information,Bueno2019-zh}.  To compute the unitary complexity, we choose the generators as the Heisenberg algebra $\{x,p,i\hbar\mathbb{I}\}$, and consider the unitary (\ref{eq:targetu}) given by a generic initial perturbation, 
\begin{equation}
    U_\epsilon = e^{-iHt} e^{i\epsilon_1 \hat{x} + i\epsilon_2 \hat{p}} e^{iHt}.
\end{equation}
One can leave out the identity generator, as it amounts to a global phase, along which we assume that the cost function vanishes. This target unitary can be solved exactly, i.e.,
\begin{equation}\label{eq:IHOunitary}
   U_\epsilon = e^{i\epsilon_1(t)\hat{x}+i\epsilon_2(t)\hat{p}},
\end{equation}
where
\begin{equation*}
\begin{aligned}
\epsilon_{1}(t)&=\epsilon_{1} \cosh(\Omega t)+\epsilon_{2}\sinh(\Omega t)/\Omega,\\
\epsilon_{2}(t)&= \epsilon_{1} \sinh(\Omega t)\Omega + \epsilon_{2} \cosh(\Omega t).
\end{aligned} 
\end{equation*}
Since the associated Hamiltonian is independent of the protocol time $\sigma$ over small distances, one can straightforwardly find a minimal geodesic, that is,
\begin{equation}
U_{short}(\sigma)=e^{i(\epsilon_1(t)\hat{x}+\epsilon_2(t))\hat{p})\sigma},\  \sigma \in [0,1].
\end{equation} 
The optimal path is therefore identified as
\begin{equation}\label{eq:geodisc} 
Y^{(1,2)}_{c}(\sigma)=\epsilon_{(1,2)}(t)\sigma,
\end{equation}
which gives the linear response matrix $L=R^\dag R$ by
\begin{equation}\label{eq:Rmatrix}
\hat{R}^u=
\begin{pmatrix}
\cosh(\Omega t) & \Omega\sinh(\Omega t)\\
\sinh(\Omega t)/\Omega & \cosh(\Omega t)
\end{pmatrix}.
\end{equation}
The OTOC matrix $\hat{O}$ can be computed as well~\cite{Hashimoto2017-ug,yan2019information}, e.g., $O_{x,p}\equiv[\hat{x}(t),\hat{p}]=i\cosh(\Omega t)$. It is straightforward to verify the OTOC-complexity correspondence (\ref{eq:COTOC}).

The state complexity in general, as has been discussed before, depends on the initial state and differs from the unitary complexity. However, we argue that for the case of the IHO and Heisenberg generators, the optimal path connecting the perturbed the unperturbed wavefunctions is the same as the unitary complexity case, namely, it is given by the geodesic (\ref{eq:geodisc}).   

For initial states that differ by a small generic perturbation, i.e., $|\Psi_2\rangle=e^{i\epsilon_1\hat{x}+i\epsilon_2\hat{p}}|\Psi_1\rangle$, the Heisenberg operator (\ref{eq:IHOunitary}) is one particular unitary protocol that connects them at any time, despite the fact that it is not the unique one. The effect of this operator is to shift the wavefunction for a fixed amount in the coordinate space, i.e., $e^{i\epsilon_2(t)\hat{p}}$ transfers a wavefunction $\psi(x)$ in the position representation to $\psi(x+\epsilon_2(t))$. We thus conclude that the two time-dependent wavefunctions have Wigner representations related via a coordinate transformation $x \rightarrow x+\epsilon_2(t)$ and $p \rightarrow p+\epsilon_1(t)$. The global phase of the wavefunction is ignored since we can generate it with the identity operator at zero cost. This coordinate transformation fixes the boundary condition for the protocol paths. One can then identify the optimal one as the straight line in the coordinate space, which is given by the geodesic (\ref{eq:geodisc}). 

The state complexity linear response matrix thus equals the one in the unitary case, from which we can extract the Lyapunov spectrum (\ref{eq:Lspectrum}) as $\{\pm \Omega\}$. This exactly matches the classical one.

\vspace{5pt}
To summarize, the novel distance measure for wavefunctions induced by the relative complexity has been applied to study chaotic signatures in quantum systems. We have developed a linear response theory that resolves the fine structure of the complexity responses to perturbations. This turns out to be the correct framework to generalize classical chaos in the phase space, i.e., the spectrum of the complexity response matrix recovers the classical Lyapunov spectrum. Moreover, we have shown that the complexity linear response is intrinsically related to the out-of-time order correlators.

\begin{acknowledgements}
The authors would like to thank Adam Brown,  Javier M. Magan, Pablo Bueno, Hugo Marrochio and Nikolai Sinitsyn for valuable feedback and inspiring discussions. We are particularly indebted to  Wojciech H. Zurek for careful readings of the manuscript and stimulating discussions, especially for suggesting the use of the Wigner function. B.Y. was supported by the U.S. Department of Energy, Office of Science, Basic Energy Sciences, Materials Sciences and Engineering Division, Condensed Matter Theory Program. B.Y. also acknowledges partial support from CNLS. W.A.C acknowledges support provided by the Institute for Quantum Information and Matter, Caltech, and for the stimulating environment from which the author had been significantly benefited. W.A.C gratefully acknowledges the support of the Natural Sciences and Engineering Research Council of Canada (NSERC).
\end{acknowledgements}

\bibliography{reference}

\end{document}